\documentclass[letterpaper]{aipproc}
\layoutstyle{8x11single}

\begin{document}
\title{Jet Modification in a Brick of QGP Matter}
\author{C.E.Coleman-Smith, G-Y.Qin, S.A.Bass \& B.M\"uller }{
  address={Physics Department, Duke University, NC, 27701},
  email={cec24@phy.duke.edu}}


\begin{abstract}

We have implemented the LPM effect into a microscopic transport model with partonic degrees of freedom  by following the algorithm of Zapp \& Wiedemann. The Landau-Pomeranchuk-Migdal (LPM) effect is a quantum interference process that modifies the emission of radiation in the presence of a dense medium. In QCD this results in a quadratic length dependence for radiative energy loss. This is an important effect for the modification of jets by their passage through the QGP. 

We verify the leading parton energy loss in the model against the leading order Baier-Dokshitzer-Mueller-Peigne-Schiff-Zakharov (BDMPS-Z) result. 

We apply our model to the recent observations of the modification of di-jets at the LHC. 

\end{abstract}

\maketitle


Jets are suppressed in heavy ion collisions. Observations of the suppression of high-$p_{T}$ hadrons \cite{PhysRevLett.89.202301, PhysRevLett.88.022301}, di-hadron correlations \cite{PhysRevLett.90.082302, PhysRevLett.91.072304} and the non suppression of photons at RHIC suggest that this is a final state partonic effect. This implies the existence of a dense colored medium through which the jets are passing. For a review of jet energy loss see \cite{Majumder:2010qh}. 

The recent observations of highly asymmetric di-jet pairs made by the ATLAS \cite{Aad:2010bu} and CMS \cite{Chatrchyan:2011sx} collaborations usher in a new and exciting era of jet tomography. 

Parton cascades describe the full space-time evolution of a system of quarks and gluons using pQCD cross sections. Their equal treatment of jet and medium makes them well suited for describing the production and propagation of jets in a hot and dense QGP matter.

The LPM effect \cite{LandauPom, PhysRev.103.1811} is a  QCD coherence effect arising from the interference of radiated quanta with medium quanta. This interference process leads to a characteristic radiation induced energy loss for the parent parton, to leading order $\Delta E \sim L^2$  \cite{ Baier:1996kr, Zakharov:1997uu}. We have modified the VNI/BMS  \cite{Bass:2002fh, Geiger:1991nj} parton transport code to approximate this effect through the introduction of dynamically modified formation times describing the finite duration of the radiation process  \cite{Zapp:2008af, ColemanSmith:2011wd}. 

In this proceeding we present an overview of the implementation of the LPM effect in a Monte-Carlo code, validate our model against the BDMPS-Z result and compute the LHC di-jet asymmetry in a simple model of a nuclear collision.


The VNI/BMS parton cascade model (PCM) is a Monte-Carlo model of the relativistic Boltzmann transport of quarks and gluons
\begin{equation}
\label{eqn-boltzmann}
p^{\mu} \frac{\partial}{\partial x^{\mu}} F_k(x, p) = \sum_{i}\mathcal{C}_i F_k(x,p).
\end{equation}
The collision term $\mathcal{C}_i$ includes all possible $2\to2$ interactions and final-state radiation $2 \to n$
\begin{equation}
  \label{eqn-pcm-collision}
  \mathcal{C}_i F_k(x,\vec{p}) = \frac{(2 \pi)^4}{2 S_i} \int \prod_{j} d\Gamma_j | \mathcal{M}_i | ^2 \delta^4\left(P_{in} - P_{out}\right) D(F_k(x, \vec{p})),
\end{equation}
$d\Gamma_j$ is the Lorentz invariant phase space for the process $j$, $D$ is the collision flux factor and $S_i$ is a process dependent normalization factor. A geometric interpretation of the total cross-section is used to select collision pairs. Between collisions, the partons propagate on-shell along classical trajectories. 

A QGP medium is simulated as a periodic box of quarks and gluons in thermal equilibrium generated at a fixed temperature, typically $T = 0.35$ GeV. 


Since the PCM is a Monte-Carlo process, it is difficult to naturally include interference processes such as the LPM effect. We implement an approximation using a local Monte-Carlo routine in the style of Zapp and Wiedemann \cite{Zapp:2008af}, this reproduces the leading BDMPS-Z \cite{Baier:1996kr, Zakharov:1997uu} result for light-parton energy loss in a QGP medium $\Delta E \sim L^2$. This method requires no artificial parameterization of the radiative process, it is a purely probabilistic medium-induced modification.

Radiation in QCD is not instantaneous. There is a time period during which the radiated quanta and the radiator cannot be resolved. If this process takes place in a dense medium this composite system may interact with the medium, leading to a modification of the separation process. The formation time 
\begin{equation}
  \label{eqn-formation-time-first}
  \tau_f = \frac{E}{Q} \frac{1}{Q} \approx \frac{\omega}{\mathbf{k}_{\perp}^2},
\end{equation}
is the lifetime of the virtual state and serves to characterize the emission process.

In the PCM a set of partons is produced by time-like branching after an inelastic scattering. The hardest radiated gluon is selected to represent this shower (the probe gluon) and is allowed to re-interact elastically with the medium. The rest of the partons involved in the radiation process propagate but are not allowed to interact with any other partons until their formation time expires. 

The probe parton is allowed to propagate through the medium and interact elastically with other partons in the medium, after the n-th scattering its formation time is
 \begin{equation}
   \label{eqn-formation-time-recalc}
   \tau_f^{n} = \frac{\omega}{\left(\mathbf{k}_{\perp} + \sum_{i=1}^{n} \mathbf{q}_{\perp,i}\right)^2}. 
 \end{equation}
This builds a dynamic length dependence into the evolution of the radiation process.

\begin{figure}
  \begin{tabular}{l r}
    \includegraphics[width=0.45\textwidth]{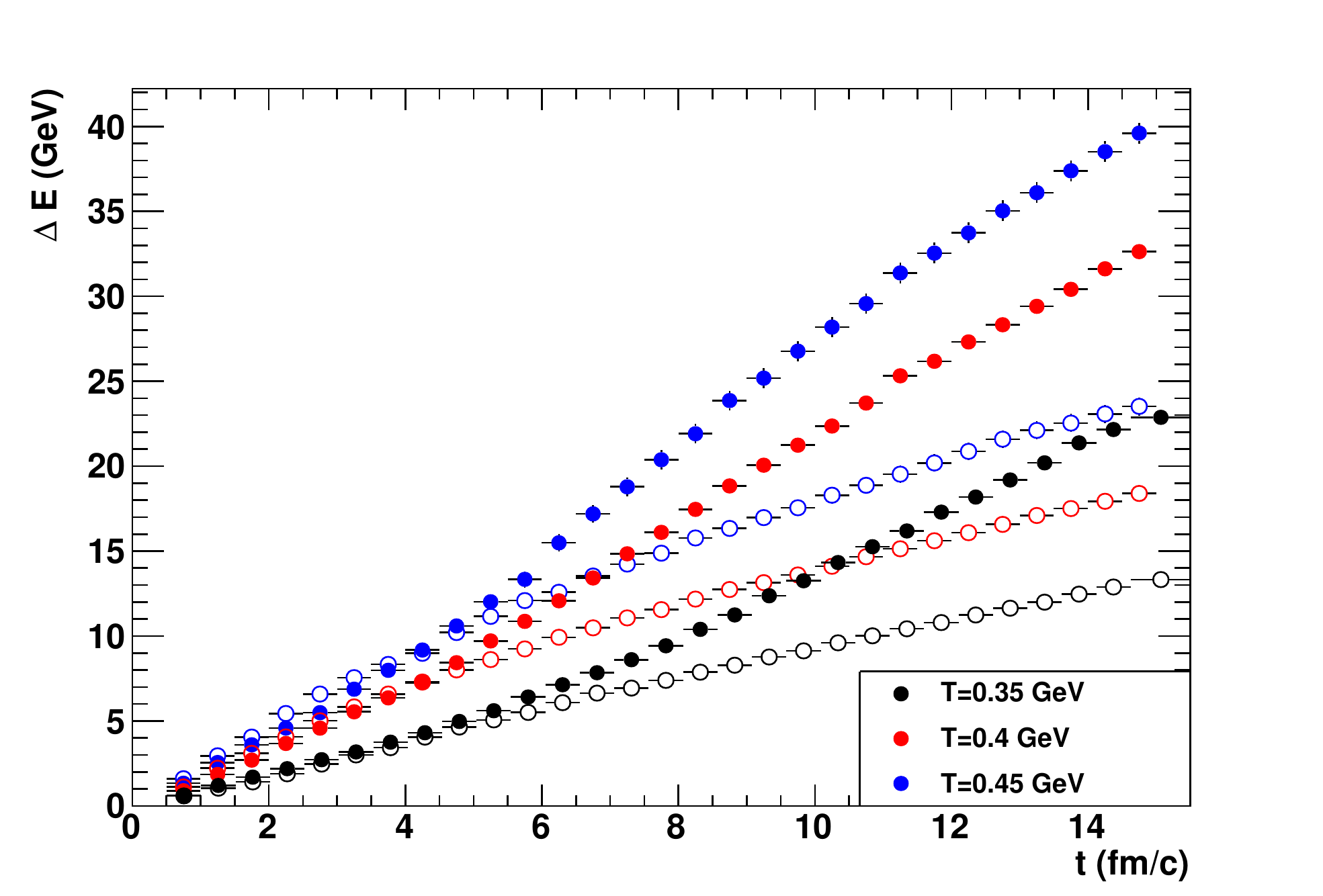} &
    \includegraphics[width=0.45\textwidth]{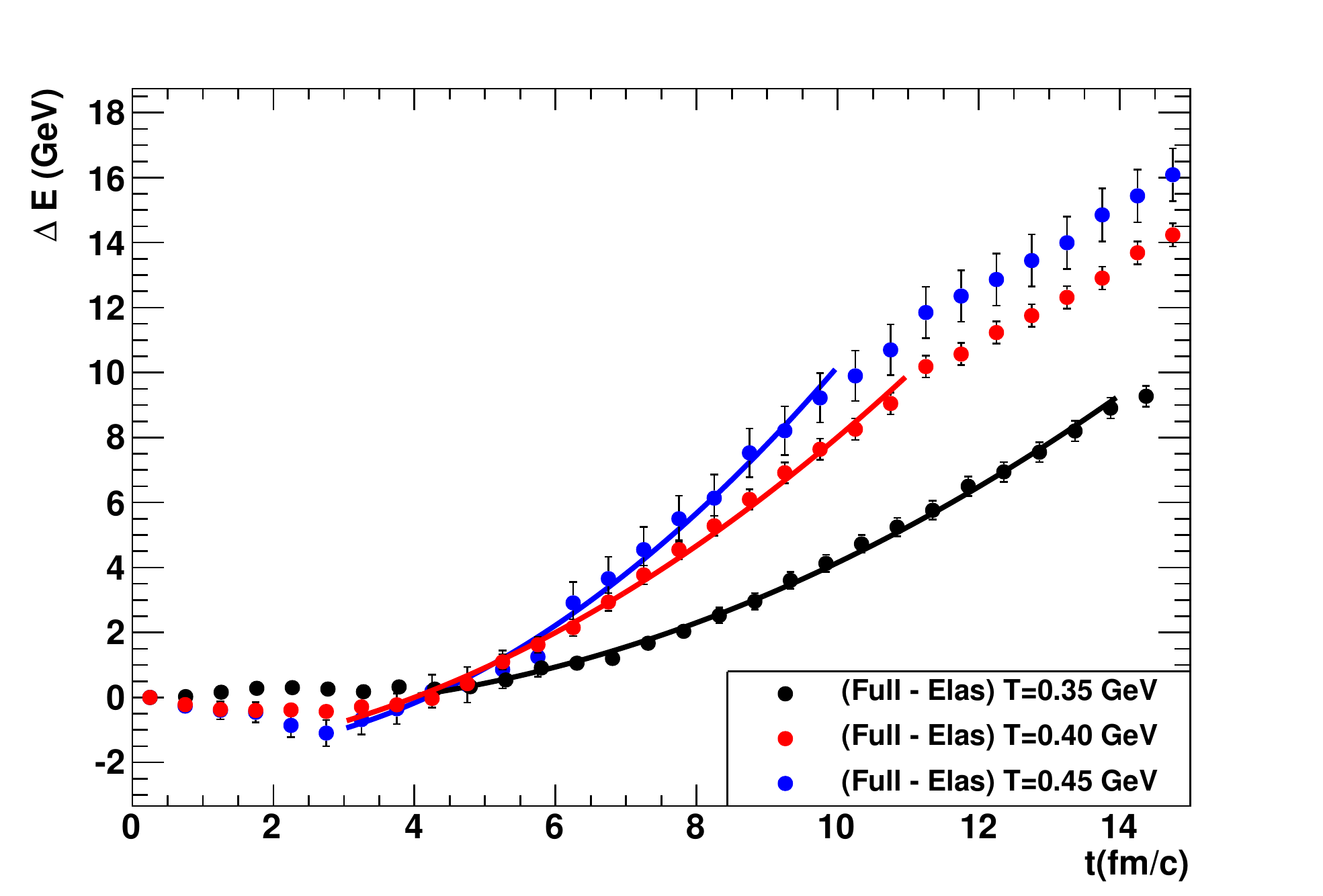}
  \end{tabular}
    \caption{Left: the total energy loss for the leading parton in a $100$~GeV jet, closed circles show the full simulation (elastic \& radiative energy loss) open circles show the energy loss for elastic scattering only. Right: the radiative only energy loss computed from the difference of the full and elastic only processes  for the leading parton in a $100$~GeV jet (closed circles). The solid lines are a fit of the form: $k_0 + k_1 t^2 \log(k_2 t)$.}
  \label{fig:detotal}
\end{figure}

We verified the performance of this proscription by fitting the radiative energy loss of the leading particle in a jet to the leading-order BDMPS-Z result \cite{Baier:1996kr}
\begin{equation}
  \label{eqn-bdmps-z}
  -\Delta E = \frac{\alpha_s C_{R}}{8} \frac{\mu^{2}}{\lambda_g} L^2 \log \frac{L}{\lambda_g}.
\end{equation}
Fig:~\ref{fig:detotal} shows the results of this analysis, by fitting with the functional form $E = k_0 + k_1 t^{2} \log(k_2 t)$ we reproduce the leading coefficient in (\ref{eqn-bdmps-z}) to less than $20\%$ over a range of medium temperatures. The elastic energy loss processes in the model are also well understood \cite{Shin:2010hu}. The VNI/BMS parton cascade is capable of accurately describing both the elastic and radiative energy loss of the leading partons in jets. 

\begin{figure}
  \begin{tabular}{l r}
    \includegraphics[width=0.45\textwidth]{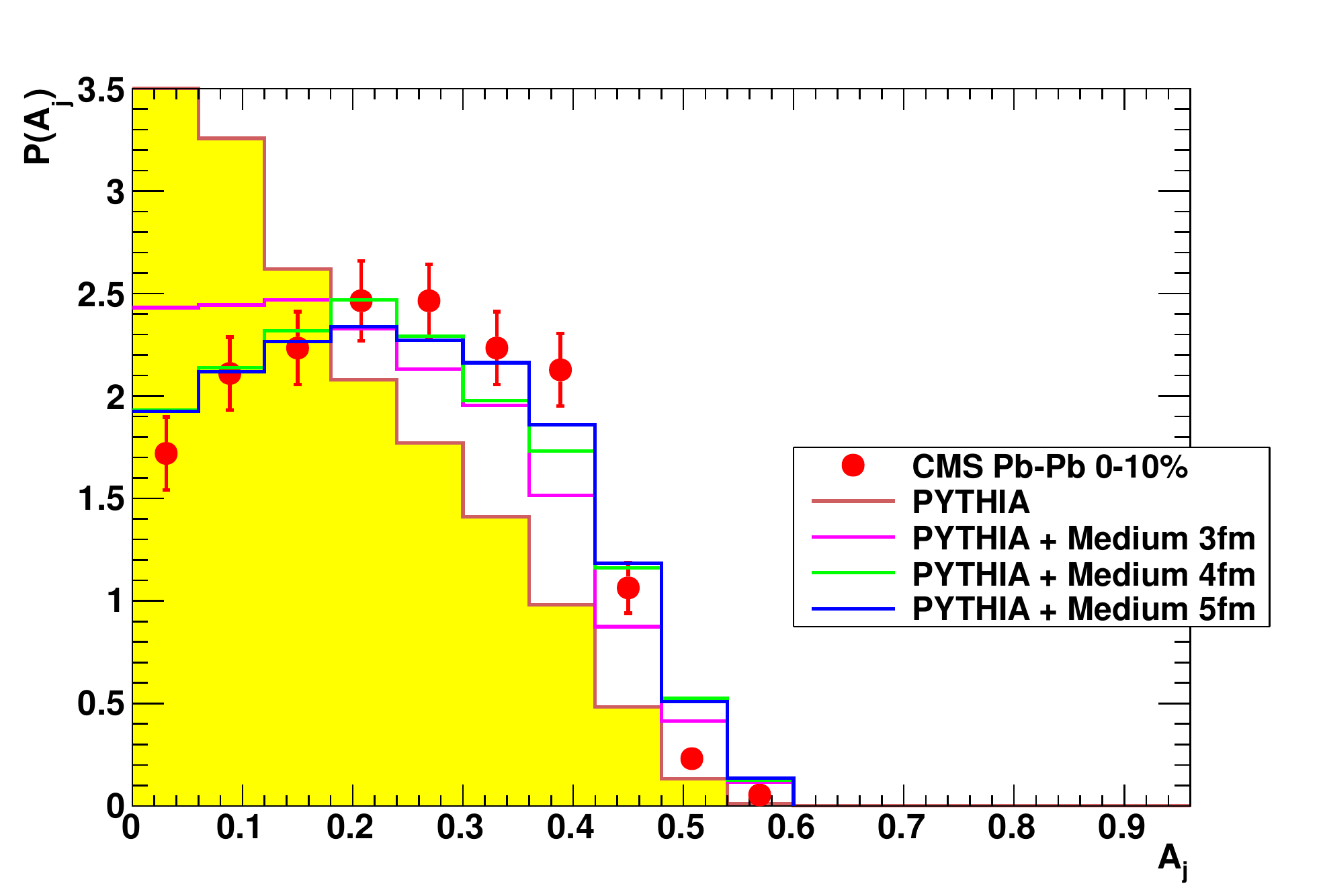} &
    \includegraphics[width=0.45\textwidth]{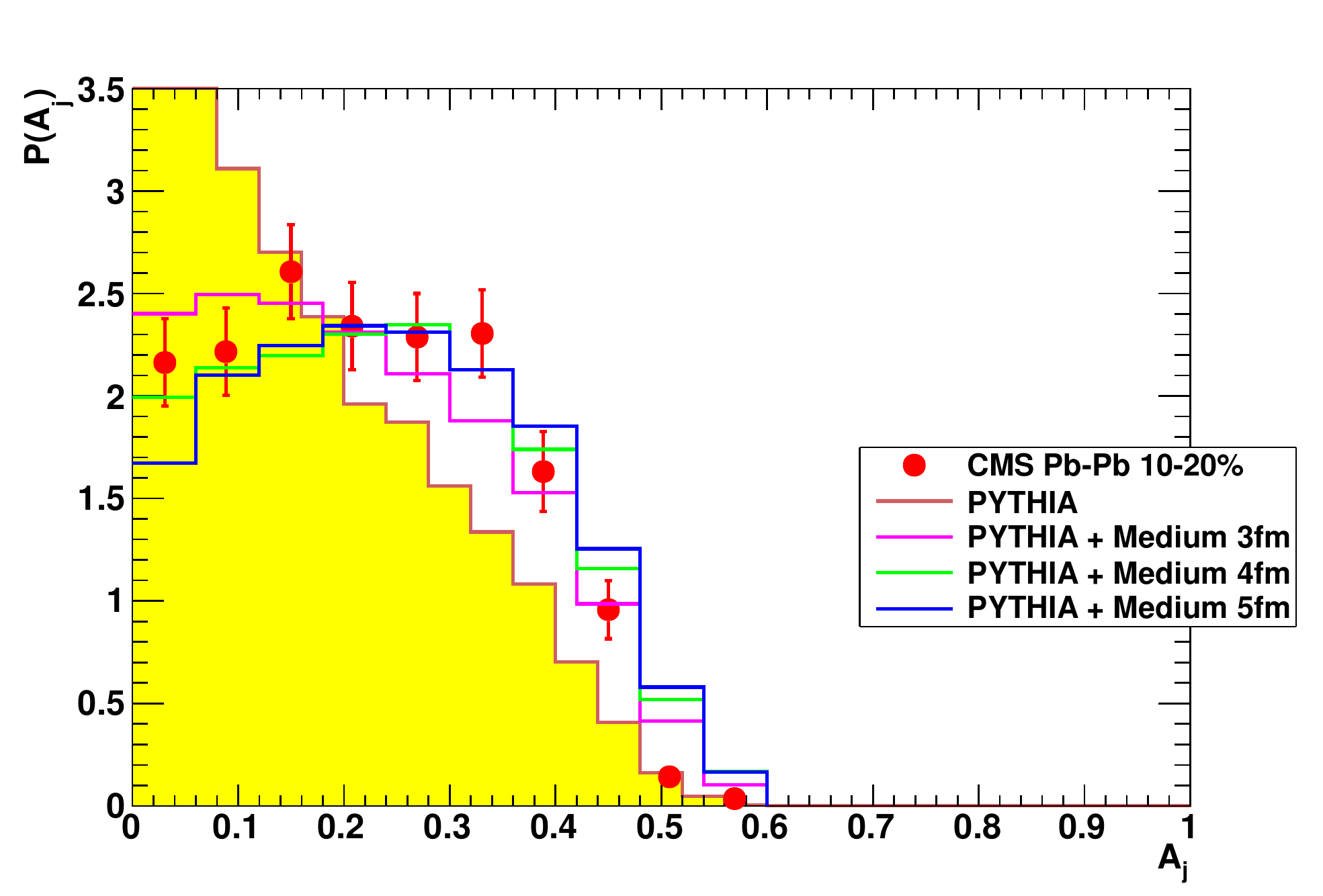}
  \end{tabular}
  \caption{Distribution of the di-jet asymmetry (red circles) $A_j$ for $0-10\%$ central (left) and $10-20\%$ central (right) collisions as measured by CMS, with $E_{t_1} \ge 120$~GeV , $E_{t_2} \ge 50$~GeV \& $\Delta \phi \ge \frac{2 \pi}{3}$. The PCM medium was fixed at $T=0.35$~GeV and $\alpha_s = 0.3$. The average medium length for the jet evolution was varied between $3-5$~fm, solid lines.}
  \label{fig:cms-aj}
\end{figure}

The di-jet asymmetry 
\begin{equation}
  \label{eqn-aj}
  A_j = \frac{E_{T,1} - E_{T,2}}{E_{T,1} + E_{T,2}},
\end{equation}
as measured by the CMS and ATLAS collaborations has shown a strong modification of sub-leading jets in a di-jet pair compared to jets in PP. In Fig:~\ref{fig:cms-aj} we compare the CMS measurement at two centralities against the PCM. 

We generate candidate di-jet events from Monte-Carlo PP collision data using PYTHIA, satisfactory jets were identified by an  anti-kt jet-finder  \cite{Cacciari:2008gp} with $R = 0.4 = \sqrt{\Delta \phi^2 + \Delta \eta ^2}$. The constituent partons of each identified jet were translated into the parton cascade medium and evolved. 

In the PCM jet paths  are sampled from a simple geometric model of the QGP, the temperature of the brick is held constant throughout the evolution of the jet. We are able to reproduce the measured results with an average path length of $R \sim 5$~fm for $0-10\%$ centrality and $R \sim 3$~fm for $10-20\%$ centrality. 

The power of our model lies in its simplicity, the parton level energy loss processes are sufficient to study jet modification without the additional complications induced by a more realistic treatment of the QGP. The dependence on the underlying energy loss process can be readily examined within our simple model framework without the complications of flow, initial state fluctuations and jet reconstruction. In a future paper we will present the sensitivity of jet energy loss, as examined through the di-jet asymmetry, to the variation of medium temperature, color screening length, strong coupling constant and in medium path length. 

We acknowledge support by DOE grants DE-FG02-05ER41367 and DE-SC0005396. This research was carried out using resources provided by the Open Science Grid, which is supported by the DOE and the NSF.

\bibliographystyle{aipproc}

\end{document}